\documentstyle[epsfig]{aipproc}

\begin{document}

\title{n-p Short-Range Correlations \\ from (p,2p + n) Measurements}

\author{A. Tang$^a$, \underline{J. W. Watson}$^a$,
J. Alster$^b$, G. Arsyan$^{d,c}$, Y. Averichev$^{i}$, \\
D. Barton$^d$, V. Baturin$^{f,e}$, N. Bukhtoyarova$^{d,e}$, A. Carroll$^d$, \\
S. Heppelmann$^f$, T. Kawabata$^g$, A. Leksanov$^f$, Y. Makdisi$^d$, \\
A. Malki$^b$, E. Minina$^f$, I. Navon$^b$, H. Nicholson$^h$, A. Ogawa$^f$, \\
Yu. Panebratsev$^{i}$, E. Piasetzky$^b$, A. Schetkovsky$^{f,e}$,
\\S. Shimanskiy$^{i}$, H. Yoshida$^g$, D. Zhalov$^f$}

\address{$^a$Dept. of Physics, Kent State Univ., Kent, OH 44242, U.S.A.\\
$^b$School of Physics and Astronomy, Sackler Faculty of Exact 
  Sciences, Tel Aviv University, Ramat Aviv 69978, Israel \\
$^c$Yerevan Physics Institute, Yerevan 375036, Armenia\\
$^d$Collider-Accelerator Department, Brookhaven National Laboratory, Upton, NY 11973, USA\\
$^e$Petersburg Nuclear Physics Institute, Gatchina, St. Petersburg 188350, Russia \\
$^f$Physics Department, Pennsylvania State University, University Park,
 PA 16801, U.S.A.\\
$^g$Dept. of Physics, Kyoto Univ., Sakyoku, Kyoto, 606-8502, Japan\\
$^h$Dept. of Physics, Mount Holyoke College, South Hadley, MA 01075, U.S.A.\\
$^i$J.I.N.R., Dubna, Moscow 141980, Russia}

\maketitle

\begin{abstract}
Recently, a new technique for measuring short-range NN
correlations in nuclei (NN SRCs) was reported by the E850
collaboration, using data from the EVA spectrometer at the AGS at
Brookhaven Nat. Lab.  In this talk, we will report on a larger set of
data from new measurement by the collaboration, utilizing the same technique.
This technique is based on a very simple kinematic approach.  For
quasi-elastic knockout of protons from a nucleus  ($^{12}$C(p,2p) was used 
for the current work), we can reconstruct the momentum {\bf p$_f$}
of the struck proton in the nucleus before the reaction, from the three 
momenta of the two detected protons, {\bf p$_1$} and {\bf p$_2$} and 
the three momentum
of the incident proton, {\bf p$_0$} :
\begin{center}
{\bf p$_f$} = {\bf p$_1$} + {\bf p$_2$} - {\bf p$_0$}
\end{center}
If there are significant n-p SRCs, then we would expect to find a neutron
with momentum -{\bf p$_f$} in coincidence with the two protons, provided 
{\bf p$_f$} is larger than the Fermi momentum $k_F$ for the nucleus
(${\sim}$220 MeV/c for $^{12}$C). Our results reported here confirm 
the earlier results from the E850 collaboration.
\end{abstract}

\section*{}

For the past half century the dominant model for the structure of nuclei, 
especially light nuclei, has been the nuclear shell model. In the shell 
model, the long-range (${\sim}$ 2 fm) part of the N-N force, in combination 
with the Pauli principle, produces an average potential in which the 
nucleons undergo nearly independent motion, and the residual interactions 
can be treated by perturbation theory. However the N-N interaction is 
also highly repulsive at short-range (${\sim}$ 0.4 fm) and it has long been 
a goal of nuclear physics to observe the effects of this short-range 
repulsion. These effects are most easily pictured in terms of momentum 
correlations rather than in terms of spatial correlations. When 
two nucleons in a nucleus interact at short range, they must have large 
relative momenta (because of their strong repulsion at short range). 
Following such a collision they will have equal and opposite momenta 
in their two-body c.m. frame. Typically, to obtain high enough relative 
momenta to probe the N-N repulsive core, one would expect the 
two-body c.m. frame to coincide roughly with the c.m. frame for the 
nucleus as a whole.

Recently, Aclander et al.[1] described new technique for observing 
such short-range correlations using data taken with the EVA 
spectrometer[2,3] at the AGS. This technique is based on a very simple 
kinematic approach. For the quasi-elastic knockout of protons from 
nuclei, e.g. $^{12}$C(p,2p) in [1] and for this work, we can 
reconstruct (event by event) the three momentum {\bf p$_f$} that each 
struck proton had before the reaction:
\begin{equation}
   {\bf p_{\em f}} = {\bf p_{1}} + {\bf p_{2}} - {\bf p_{0}}
\end{equation}
where {\bf p$_0$} is the momentum of the incident proton and 
{\bf p$_1$} and {\bf p$_2$} are the momenta of the two detected 
protons. The question we then ask is whether or not there is 
a coincident neutron with {\bf p$_n$} $\approx$ -{\bf p$_f$}. To answer 
this question, we deployed 36 neutron detectors to look for 
triple coincidences of the kind $^{12}$C(p,2p+n).

The EVA spectrometer is designed to detect proton pairs from quasielastic 
collisions with ${\theta}_{cm}$  $\approx$ 90$^\circ$. Because the 
cross section 
for this geometry fall steeply with the Mandelstam variable s, the (p,2p) 
reaction preferentially occurs for nuclear protons with {\bf p$_f$} in the 
forward going lab direction. Therefore most of our 36 neutron detectors 
were placed in the backward laboratory hemisphere.  Figure 1 shows the 
layout of the 36 neutron detectors relative to EVA. The detectors 
in arrays 1 and 2 had dimensions 10 cm x 12.5 cm x 1m. The detectors 
in array 3 had dimensions 25 cm x 10 cm x 1 m.

Our initial objective with the new, more extensive, triple coincidence
measurement with EVA in 1998 was to confirm the results from the 1994 
data reported in [1]. To this end we applied the following cuts:
\newcounter{bean}
\begin{list}%
{\ \ \ \ \ \ \ (\arabic{bean}).}{\usecounter{bean} \setlength{\rightmargin}{\leftmargin}}
 \item  There should be two (and only two) high {\bf p$_t$} positive tracks - 
kinematics dictates that these are both protons.
 \item  The missing energy, E$_{miss}$ should be appropriate for 
quasi-elastic scattering (within our resolution): 
	-0.2 $<$ E$_{miss}$ $<$ 0.8 GeV.
 \item  0.05 $<$ {\bf p$_n$} $<$ 0.55 GeV/c.
%
\end{list}

There is another cut we can apply to more fully reproduce the conditions 
of the 1994 run. For that data, only the straw-tube sectors near the 
midplane of EVA were working. So we can impose a fourth cut:
\begin{list}%
{\ \ \ \ \ \ \ (4).}{\usecounter{bean} \setlength{\rightmargin}{\leftmargin}}
 \item  The two detected protons were limited to a plane parallel to 
the neutron detectors within $\pm$ $25^\circ$.
\end{list}

Figure 2 shows our preliminary momentum correlation results for 
$^{12}$C(p,2p+n) at 5.9 GeV/c with cuts (1), (2), (3) and (4). 
Figure 2 is a plot of {\bf p$_{fx}$}, the reconstructed 
x component of {\bf p$_f$}, vs. the measured {\bf p$_n$}.
Since the neutrons are detected largely going ``downward'' in the 
laboratory we would expect {\bf p$_{fx}$} to be ``upward'' for correlated 
high-momentum n-p pairs. We see in Fig. 2 that for events with 
{\bf p$_n$} $>$ 0.22 GeV/c (the Fermi momentum for $^{12}C$) 
that indeed {\bf p$_{fx}$} is predominantly ``upward''. 
For {\bf p$_n$} $<$ 0.22 GeV/c there is no evident correlation, which is 
also as expected. For {\bf p$_n$} $\geq$ 0.22 GeV/c, the ratio of events with 
{\bf p$_{fx}$} $\geq$ 0 and {\bf p$_{fx}$} $<$ 0 in 
Fig. 2 is 18/3. This result is completely consistent with the ratio of 
17/1 reported in [1] and provides strong confirmation of that result.

With  cut (4) we are selectively rejecting events with 
an approximately ``coplanar'' geometry where all four transverse 
momenta {\bf p$_{t1}$}, {\bf p$_{t2}$}, {\bf p$_{tf}$} and {\bf p$_{tn}$} 
all lie roughly in the same plane with 
$\Delta \phi$ = $|$$\phi{_2}$ - $\phi{_1}$$|$ $\approx$ 180$^\circ$.
The type of events selected by (4) are then of the ``non 
co-planar'' type where $\Delta \phi$ differs 
significantly from 180$^\circ$. This preferential selection of non co-planar 
events has an unintended benefit, in that our 
reconstruction of {\bf p$_f$} has much better resolution for non co-planar 
events than for co-planar events. 

The data in Figs. 2 represents 10\% to 20\% of the total data recorded 
in 1998. As we continue the analysis, we will be exploring the 
kinematic and geometric constraints on where we find strong evidence 
of n-p correlations - such as were reported in [1] and are seen in Fig. 2.

This research was supported by the U.S. - Israel Binational Science 
foundation, the Israel Science Foundation founded by the Israel 
Academy of Sciences and Humanities, the NSF grants PHY-9501114, 
PHY-9722519 and U.S. Department of Energy grant DEFG0290ER40553.

\begin{figure}[b!] 
\centerline{\epsfig{file=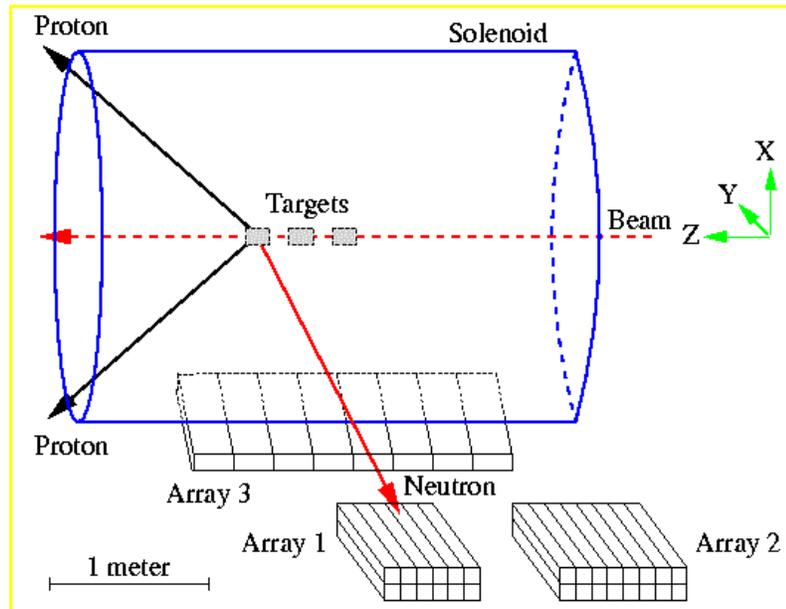,height=3.7in,width=4.5in}}
\vspace{5pt}
\caption{Layout of the experiment}
\label{foobar:fig1}
\end{figure}

\begin{figure}[b!] 
\epsfxsize=3.1in
\epsfysize=3.1in
\epsffile[-180 150 320 650]{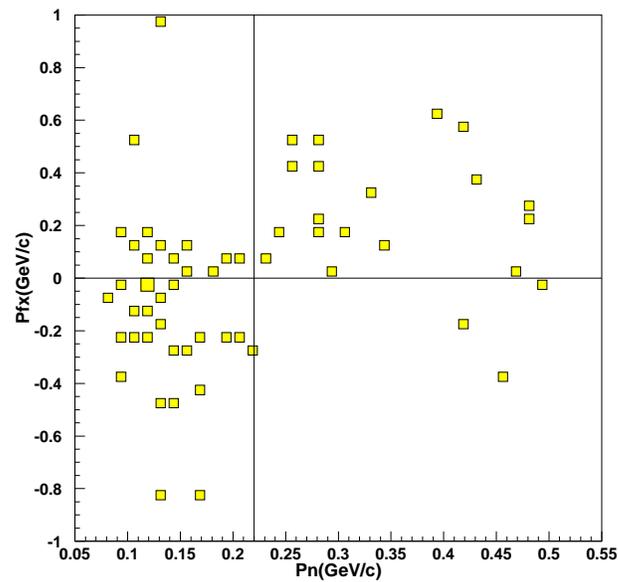}
\vspace{5pt}
\caption{Pfx vs. Pn with cuts 1,2,3 and 4 for $^{12}$C(p,2p+n) at 5.9 GeV/c}
\label{foobar:fig2}
\end{figure}

\end{document}